\documentclass{article}
\usepackage{graphicx} 

\title{Alternate Learning and Compression approaching $R(D)$}
\author{Ram Zamir (TAU) and Kenneth Rose (UCSB)}
\date{May 2024}

\begin{document}

\maketitle



\section*{Extended Abstract} 
\subsubsection*{Presented at `Learn 2 Compress', workshop at ISIT 2024, Athens}

The inherent trade-off in on-line learning is between exploration and exploitation. A good balance between these two (conflicting) goals can achieve a better long-term performance. Can we define an optimal balance? We propose to study this question through a backward-adaptive lossy compression system, which exhibits a ``natural'' trade-off between exploration and exploitation.


From a rate-distortion perspective, a universal lossy compression system needs to learn the rate-distortion achieving distribution, $Q^*$, in order to construct an optimal codebook. 
This is unlike the lossless case, where strings of the source itself have the ``right'' distribution and can serve as a basis for the code dictionary (as done, e.g., in Tunstall or Lempel Ziv coding \cite{LZ2}). 
When the source distribution is known, the system designer can use the well known Arimoto-Blahut algorithm to compute $Q^*$
\cite{Arimoto72,Blahut72}. 
Less known is the fact that even if the source distribution is unknown, a sequential scheme can learn $Q^*$ on the fly, while compressing consecutive source vectors. 

Such an asymptotically optimal sequential coding scheme, called Natural Type Selection (NTS), was described in \cite{NTS_original}. 
It is based on the observation that the empirical distribution of the $d$-matching codeword is atypical with respect to, and better than, the codebook generating distribution.
Hence, this empirical distribution provides information on how to adjust the codebook generating distribution in the direction of $Q^*$. 

Specifically, NTS-based encoding alternates between two phases:
\begin{enumerate}
    \item {\em Compression phase} - where the encoder finds the index of the first $d$-matching codeword and transmits it to the decoder; 
and 
  \item {\em Learning phase} - where the encoder and decoder estimate the type (or some other representative parameter) of the $d$-matching codeword.
\end{enumerate}
After these two phases, the codebook is updated (identically) by the encoder and the decoder.
As it turns out, \cite{NTS_original}, in the limit of a large word length $L$,
the compression and learning phases above stochastically simulate an iteration of the Blahut algorithm for RDF computation. 
Hence, after many such iterations NTS converges to the RDF; see Arimoto \cite{Arimoto72} and Blahut \cite{Blahut72}.\footnote{This description amounts to a fixed-distortion variant of Blahut \cite{Thomas_Cover,Alternating_Min}. 
Replacing $d$-match by a weighted distortion-code-length sum amounts to the usual ``fixed-slope'' version of the Blahut algorithm \cite{Blahut72}.}

The NTS mechanism was extended and investigated from various information theoretic perspectives and settings,
\cite{NTS_parametric,NTS_ISIT,NTS_memory,Hila-Rami2023,correct-mistakes2006,kochman2003computation,TridenskiZamir2020}.
In this extended abstract we propose to consider its implications within the context of on-line learning and reinforcement learning, e.g. \cite{Exploration-Exploitation2002}.
This is a preliminary study, and we do not formally establish new results.  
Rather, we propose a fresh look that we believe may be of some interest to researchers in the intersection of compression and learning.


\subsection*{Why we need to explore?}

Exploration is {\em not} intrinsically inherent to universal compression, but rather a consequence of a {\em backward}-adaptive model of such system.
Let us clarify the difference between the two modes of adaptation: forward and backward.
In forward (``batch'' / ``two-part'') adaptation, e.g., dynamic Huffman or CELP speech coding, the encoder learns the source statistics, computes the optimal encoding parameters, and sends them to the decoder as a header (``side information''), before it begins to encode the source data.  
In backward (``sequential'') adaptation, e.g., Lempel-Ziv or ADPCM, both the encoder and decoder learn the parameters from past reconstructed samples, so there is no explicit transmission of side information.
See 
\cite{jayant1984digital,gersho2012vector,gibson1998digital}.
While in lossless compression the two modes of operation are essentially equivalent (e.g. \cite{willems1997reflections}), 
in the lossy case they are fundamentally different:
forward adaptation learns from the {\em clean} version of the source, while backward adaptation learns from the {\em noisy} (quantized) version.
Furthermore, the difference between the two grows in significance with increase in the prescribed distortion level. 
In information-theoretic terms, as $D$ increases from $0$ to $D_{\rm max}$, the reconstruction distribution $Q^*$, which achieves the rate-distortion bound, deviates from the source distribution $P$, progressively concentrates on a smaller subset of the reconstruction alphabet, and eventually at $D=D_{\rm max}$ collapses to a single probability mass point at the ``centroid'' letter.

We argue that for a memoryless source and a given (mismatched) reconstruction codebook, 
the type $Q$ of the reconstruction sequence is a {\em sufficient statistic} for learning $Q^*$ 
in a backward mode.
Furthermore, 
at large distortion ($D$ close to $D_{\rm max}$),
$Q$ carries almost no information about the source distribution $P$ itself, and therefore $Q^*$ cannot be computed directly from $Q$.  
Thus, a type's goodness (for compression) can only be established when a codeword of this type d-matches a source word.
We thus conclude that in backward-adaptive lossy compression at high distortion, explicit exploration of types is necessary in order to find $Q^*$.   




\subsection*{Rates of convergence}


To study the efficiency of NTS, we first consider the
speed of convergence 
of several related learning algorithms.

The convergence of the Blahut algorithm to the RDF is of the order of $O(1/N)$ after $N$ iterations. This was shown in (\cite{Boukris_convergence,Alternating_Min}) by writing the sum of the gaps of the intermediate rates $R(P,Q_N,D)$ from $R(P,D)$ as a telescopic sum that is bounded by a finite constant = the divergence between the optimum output distribution $Q^*$ and the initial output distribution $Q_0$. 
Thus, the gaps must decrease at least as fast as $O(1/N)$. 
A similar decrease is believed to hold for iterative design of $K$-level quantizers, via the Lloyd-Max algorithm (alternations between centroids computation and thresholds computation). 
%
While faster-than-Blahut computation of $R(D)$ is possible at small distortions due to the known structure of the RDF (a Lagrangian solution that assumes that all the output alphabet letters have positive probability), at high distortion most letters are inactive,
so the situation is similar to computation of the optmal 2-level quantizer.




As for the effect of the word length, universal compression schemes, lossy and lossless, are known to exhibit redundancy on the order of $O(\log(L)/L)$. 
In the lossy case, this is due to finite vector-quantization loss (``granular gain''), as well as to the cost of universality (statistical learning).  



\subsection*{Exploration strategies}

The two-phase compression-learning mechanism of NTS can be viewed as an exploitation step (compression), followed by an exploration step (learning).
The amount of exploration is governed by the frequency of atypical codewords in the random codebook.
In a codebook generated i.i.d, this frequency is known to decay exponentially with the divergence between the codeword type and the codebook generating distribution. 
This divergence also governs the steps of the corresponding Blahut algorithm,
and dictates its rate of convergence.

Is this ``natural'' trade-off between exploration and exploitation optimal?

In fact, even if the source is memoryless, a codebook distribution that is not i.i.d.\ can emphasize rare types and accelerate the movement towards $Q^*$. 
A simple example is a uniform weighting of all type classes
\cite{Thomas_Cover,ZamirRose1997}.
More generally, to obtain a richer codebook distribution,
one may use a mixture over a parametric family of distributions,
as done in the universal lossless case,
e.g., in Minimum Description Length \cite{MDL1998} or Context-Tree Weighting \cite{CTW1995,willems1997reflections,MerhavFeder1995}.

This view suggests a tradeoff between ``breadth and depth'', i.e., wide exploration versus narrow exploration: 
a richer universal mixture implies higher probability of rare types and reduced probability of typical codewords. 
This may be disadvantageous as we get closer to $Q^*$, due to the inherent cost in coding rate (especially in the non-asymptotic regime where the word length $L$ is moderate). 
Hence, there should be an optimal schedule for narrowing the richness of the universal mixture as the NTS gets closer to $Q^*$.  


\bibliographystyle{abbrv}

\bibliography{ref2}

\end{document}